\documentclass[twocolumn,showpacs,prl,amsmath,amssymb,floatfix]{revtex4}
\usepackage{graphicx}% Include figure files
\usepackage{dcolumn}% Align table columns on decimal point
\usepackage{bm}% bold math
\usepackage{textcomp}
\usepackage{epsfig}
\usepackage{color}
\newcommand{\beq}{\begin{equation}}
\newcommand{\eeq}{\end{equation}}
\newcommand{\bea}{\begin{eqnarray}}
\newcommand{\eea}{\end{eqnarray}}

\newcommand{\bary}{\begin{array}}
\newcommand{\eary}{\end{array}}
\newcommand{\benum}{\begin{enumerate}}
\newcommand{\eenum}{\end{enumerate}}
\newcommand{\bitem}{\begin{itemize}}
\newcommand{\eitem}{\end{itemize}}
\begin{document}
\title{Theoretical description of time-resolved photoemission spectroscopy: application to pump-probe experiments}

\author{J. K. Freericks$^{1}$,
H. R. Krishnamurthy$^{1,2,3}$ and Th. Pruschke$^4$}
\affiliation{$^1$Department of Physics, Georgetown University,
 37th and O Sts. NW, Washington, DC 20057, USA\\
 $^2$Centre for Condensed Matter Theory, Department of Physics, Indian Institute of Science,
 Bangalore 560012, India\\
$^3$Condensed Matter Theory Unit,
Jawaharlal Nehru Centre for Advanced Scientific Research,
Bangalore 560064, India\\
$^4$Institute for Theoretical Physics, University of G\"ottingen, Friedrich-Hund-Platz 1,
D-37077 G\"ottingen, Germany}
\date{\today}

\begin{abstract}
The theory for time-resolved photoemission spectroscopy as applied to pump-probe experiments is
developed and solved for the generic case of a strongly correlated material.  The formal development incorporates all of the nonequilibrium
effects of the pump pulse and the finite time width of the probe pulse.  While our formal development is completely general, in our
numerical illustration for the Hubbard model, we assume the pump pulse drives the electrons into a nonequilibrium configuration, which rapidly thermalizes to create a hot (quasi-equilibrium) electronic system, and we then study the effects
of windowing that arise from the finite width of the probe pulse.  We find sharp features in the spectra are broadened, particularly the quasiparticle peak of strongly correlated metals at low temperature.
\end{abstract}
\maketitle

Pump-probe, femto-second (and recently, atto-second) time-resolved photoemission spectroscopy (TR-PES), and time-resolved  angle resolved  PES (ARPES)  techniques can directly examine the excited state nonequilibrium dynamics of electrons in solids~\cite{trps-attosec}, including some strongly correlated electron systems~\cite{trps07-perfetti,Perfetti-TaS2}. In these experiments, an intense pulse of radiation ``pumps'' the system into a highly excited nonequilibrium state. After a variable time delay, the system is subject to a weak ``probe'' pulse of higher energy photons, ejecting photoelectrons which are detected with energy (and angle) resolution.

Conventional (continuous probe beam) ARPES in layered materials can be well approximated~\cite{ARPESreview} as a direct measure of the momentum and frequency dependent lesser Green's function~\cite{baym-kadanoff} of the electrons in the layers (i.e., their spectral function multiplied by the Fermi function). In more isotropic (three-dimensional) materials, the spectral function gets averaged over $k_z$, the component of the momentum perpendicular to the layers. The theoretical situation is less clear for pump-probe photoemission. The interpretation generally used is that the pump creates ``hot electrons" in quasi-equilibrium at a higher effective electronic temperature ($T_{el}$) compared to the lattice (phonons), which then cool gradually, so that the probe PES essentially measures ``equilibrium'' lesser functions at different values of $T_{el}$ for different time delays between the pump and the probe pulses. For example, recent TR-ARPES experiments~\cite{Perfetti-TaS2}  on the layered material 1T-TaS$_2$ [believed to be an unusual, charge density wave (CDW)-induced, Mott insulator,] were interpreted in this way, using (equilibrium) spectral functions from a dynamical mean-field theory (DMFT)~\cite{dmft} treatment of the 2-d Hubbard model for a range of temperatures and fillings, {\em chosen as fitting parameters}, for the different time delays used. Such an approach, while reasonable in many contexts, avoids addressing two important questions connected with (1) the nonequilibrium dynamical aspects of the experiment, and (2) the effects arising from the finite width of the probe pulse (in the time domain).

In this letter, we address these questions---the first, by developing a precise formulation for what is actually measured in the TR-PES experiments, making future calculations of the  nonequilibrium effects possible; and the second, by studying the effect of the probe pulse width, but using equilibrium spectral functions at high effective electronic temperatures as in the previous studies.

We begin by developing a theoretical treatment for what is measured in TR-PES.
We assume that the system, modeled by a quantum many-body Hamiltonian $\mathcal{H}$, is in equilibrium at a temperature $T$ before the pump is turned on. It is represented in the distant past ($t \rightarrow -\infty$) by an ensemble of the (many body) eigenstates $|{\Psi}_n\rangle $ of $\mathcal{H}$, present with the Boltzmann probability $\rho_n = \mathcal{Z}^{-1}\exp[-E_n/(k_B T)]$ where $E_n$ are the corresponding energy eigenvalues, and $\mathcal{Z}=\sum_n\exp[-E_n/(k_B T)]$ is the partition function. Turning on the pump pulse modifies $\mathcal{H}$ into a time dependent Hamiltonian $\mathcal{H}_{\rm pump}(t)$ whose precise form depends on the way one models the interaction of the pump radiation [represented by the vector potential  ${\vec{A}}_{\rm pump}(\vec{r},t)$ whose $t$ dependence includes its turning on and off] with the electrons of the system~\cite{pump-model-ham}.   Then, at time $t_0$, just before the probe is turned on, the system is represented by the ensemble of states  $ |{\Psi_n^I}(t_0)\rangle \equiv U(t_0, -\infty) |{\Psi}_n\rangle $ (with the same Boltzmann probability as above), where $U(t_2, t_1) \equiv \mathcal{T}_t \{\exp [-i \int_{t_1} ^{t_2} dt' \mathcal{H}_{\rm pump}(t')/\hbar ]\}$ is the unitary time development operator of the system in the presence of the pump radiation. Here $\mathcal{T}_t$ is the time ordering operation that handles the non-commutativity between $\mathcal{H}_{\rm pump}(t^\prime)$ at different times. $\{ |\Psi^I_n (t_0)\rangle \}$ acts as the ensemble of `initial' states for the quantum transitions generated by the time-dependent probe pulse. By explicitly incorporating the time-evolution operator of the quantum system with the pump, we include all possible nonequilibrium dynamics into the formalism.

Similarly, when the probe is turned on, the Hamiltonian is modified to ${\mathcal{H}_{\rm pump}}(t)+\mathcal{H}_{\rm probe}(t)$ to appropriately include the vector potential ${\vec{A}}_{probe}(\vec{r},t)$ arising from the radiation field of the probe pulse (it is simplest to think of the pump and probe pulses as being active at nonoverlapping times, but this is actually not a requirement). If the system is in a particular `initial' state $|\Psi^I_n (t_0)\rangle$, then, at any later time $t$ it will evolve to the `final' state
\beq
|{\Psi_n^F}(t)\rangle \equiv {\tilde {U}}(t, t_0)|{\Psi_n^I}(t_0) \rangle,
\label{final-st}
\eeq
where $ {\tilde {U}}(t, t_0) \equiv \mathcal{T}_t \exp [-i \int_{t_0} ^{t} dt^\prime \{\mathcal{H}_{\rm pump}(t^\prime)+\mathcal{H}_{\rm probe}(t^\prime) \}/\hbar]$ is the full time development operator in the presence of the probe pulse. When we have nonoverlapping pump and probe pulses, $\mathcal{H}_{\rm pump}(t^\prime)=\mathcal{H}$ for $t^\prime>t_0$. 
%(For simplicity, we have not exhibited the photon Hamiltonian, but do take it into account in the ensuing discussion.)  
After the probe pulse has been turned off, the probability that a  photoelectron is in a final state with momentum $\vec{k}_e \equiv k_e \hat{k}_e$ (in a momentum interval $dk_e$ and solid angle $d\Omega_{\hat{k}_e}$) is given by
\beq
\lim_{t\rightarrow\infty}\frac{(k_e)^2 dk_e d\Omega_{\hat{k}_e}} {(2 \pi)^3}P(t); \,P(t)\equiv \sum_{n,m} \rho_n \left |\langle{\Psi}_m ; \vec{k}_e| {\Psi_n^F}(t) \rangle\right |^2.
\label{PES-probability}
\eeq
Here, as appropriate to photoemission from an experimental sample {\em with a surface}, $|{\Psi}_m ; \vec{k}_e\rangle $ is well approximated as a direct product of the many-body eigenstate $|{\Psi}_m\rangle $ of the initial (and final) time-independent Hamiltonian $\mathcal{H}$, and a {\it one-electron scattering state} which is a free electron state of momentum $\hbar\vec{k}_e$ outside the sample.  Since the eigenstate $|\Psi_m\rangle$ in which the system is left is not determined in the experiment, and the initial state can be any one of the ensemble of initial states with probability $\rho_n$, Eq.~(\ref{PES-probability}) includes an unconstrained sum over $m$, and a sum over $n$ weighted by $\rho_n$.

\begin{figure*}[t]
\centerline{\includegraphics [width=2.46in, angle=0, clip=on]  {fig1a.eps}
\includegraphics [width=2.3in, angle=0, clip=on]  {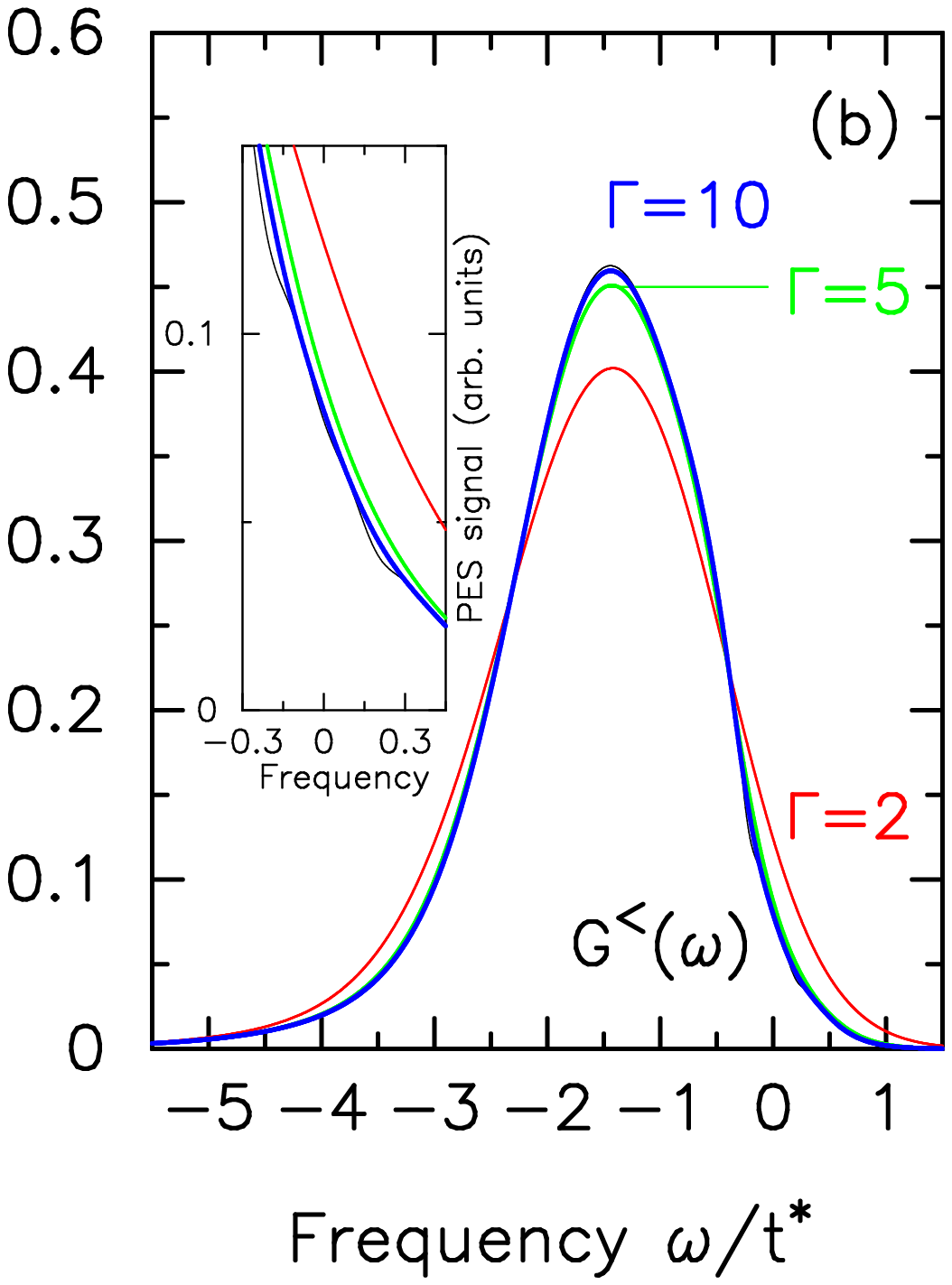}
\includegraphics [width=2.3in, angle=0, clip=on]  {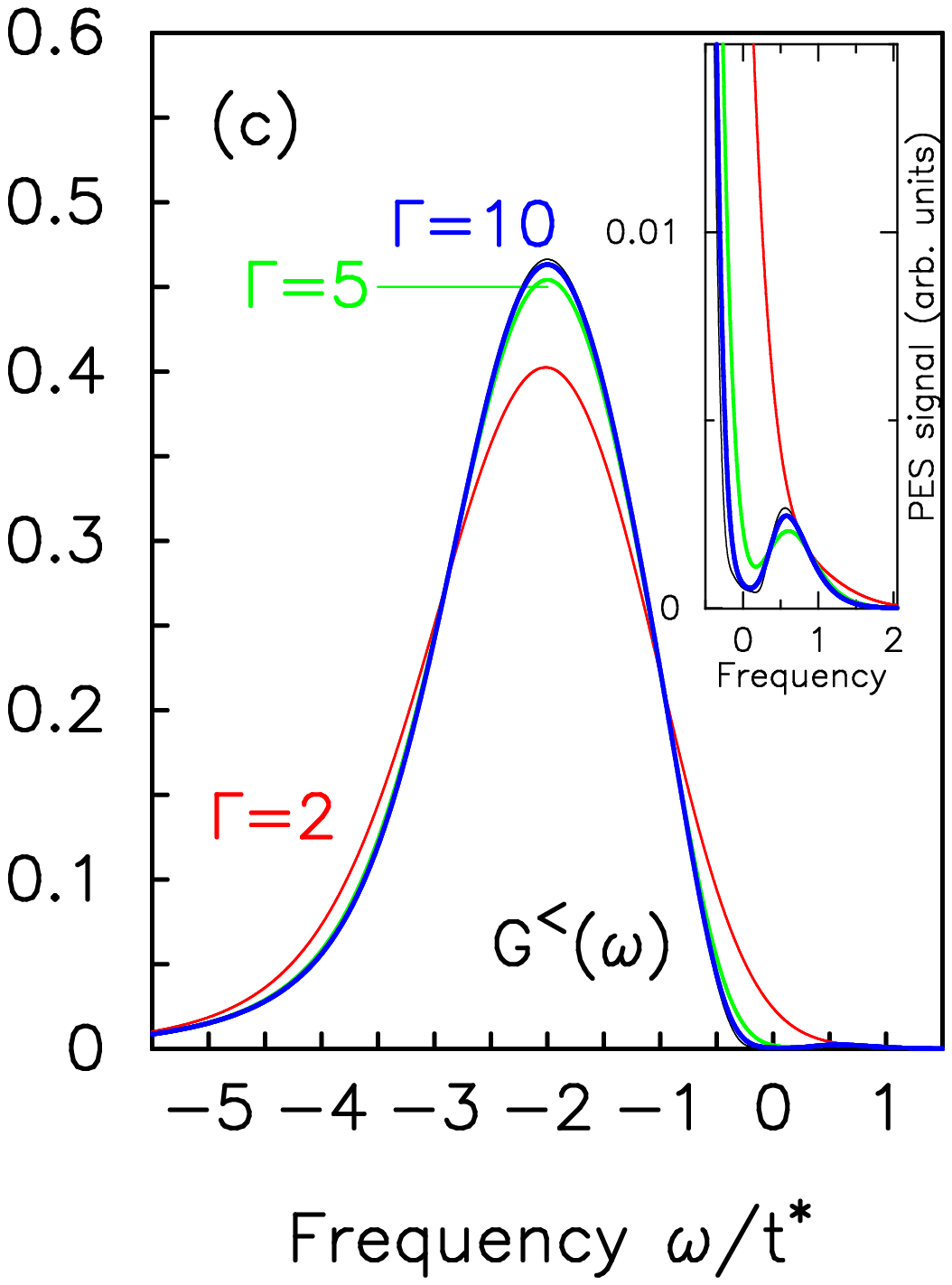}}
\caption[]{
(Color online) Photoemission spectra for the Hubbard model at half filling and $U=3$ for panels (a) and (b) (metal) and $U=4.2$ for panel (c) (insulator).  The temperatures are $T=0.00539$ for panel (a) and $T=0.0189$ for panels (b) and (c).  The continuous beam photoemission curve is in black and is labeled by $G^<(\omega)$, while the pump-probe curves for different probe widths are in other colors and labeled by their widths $\Gamma$. Note that there is very little difference between the continuous beam and $\Gamma=10$ curves in panels (b) and (c). The insets focus on the low-energy structure, where there are differences.
}
\label{fig: PES_U}
\end{figure*}

We presume that the pump pulse is intense enough that it needs to be treated non-perturbatively, but that the probe pulse is weak enough that $\mathcal{H}_{\rm probe}(t^\prime)$ can be treated by perturbation theory.  Hence, to leading order
\beq
{\tilde{U}}(t, t_0) \simeq  U(t, t_0)[1 - \frac{i}{\hbar} \int_{t_0} ^{t} dt^\prime U^{\dag}(t^\prime,t_0) \mathcal{H}_{\rm probe}(t^\prime) U(t^\prime,t_0)]
\label{U-probe}
\eeq
The pump radiation is chosen such that its photons do not have sufficient energy to overcome the work function $W$ of the sample and eject photoelectrons. Hence the photoemission process arises only from the second term in Eq.~(\ref{U-probe}). The component of $\mathcal{H}_{\rm probe}(t^\prime)$  which causes the absorption of a photon of momentum $\vec{q}$ (frequency $\omega_{\vec{q}} = cq$ ) and the ejection of an electron of momentum $\vec{k}$ inside the system as a photoelectron of momentum $\vec{k}_e$ is of the form:
\beq
\mathcal{H}_{\rm probe}(t^\prime)=\int dk_z M(\vec{q},\vec{k}, k_{ez})s(t^\prime)  a_{\vec{q}} c^{\dagger}_{\vec{k}_e} c_{\vec{k}},
\label{PES-ham}
\eeq
Here $a_{\vec{q}}$ is the annihilation operator for a photon with momentum $\vec{q}$ , and for simplicity, we have dropped the spin indices of the electron operators. The probe shape function, $s(t^\prime)$, captures the time dependence of the probe envelope, including its turning on and off. $M(\vec{q},\vec{k},k_{ez})$, the matrix element for the process, depends on the details of the modeling of the sample, especially its surface~\cite{mat-el}.
This choice assumes the sudden approximation, where the photoelectron rapidly moves out of the sample. To leading order in $\mathcal{H}_{\rm probe}$ (and using the factorization of $|{\Psi}_m ; \vec{k}_e\rangle $), we have
\begin {eqnarray}
\left |\langle {\Psi}_m ; \vec{k}_e| {\Psi_n^F}(t)\rangle \right | &\simeq&  \frac{1}{\hbar} \Big | \int dk_z M(\vec{q},\vec{k}, k_{ez})\int_{t_0} ^{t} dt^\prime s(t^\prime) e^{-i\omega t^\prime}\nonumber\\
&\times& \langle{\Psi}_m |U^{\dag}(t^\prime,t_0) c_{\vec{k}}U(t^\prime,t_0)|{\Psi_n^I}(t_0)\rangle\Big |.
\label{PES-amplitude}
\end{eqnarray}
Here $ \hbar \omega \equiv  \hbar \omega_{\vec{q}} - (\hbar k_e)^2/(2m_e) - W $  is the energy of the excitation left in the system after the photoemission process. For probe photons of fixed direction and energy, and for a given material, specifying $\omega$ determines $k_e$, hence the probability $P(t)$ in Eq.~(\ref{PES-probability}) is a function only of $t$, $\omega$ and $\hat{k}_e$. Using the properties of the time development operator, and the completeness of $\{|\Psi_m\rangle \}$  it is straightforward to show that
\bea
P(t,\omega, \hat{k}_e) & \simeq & \frac{1}{(\hbar)^2} \int dk^\prime_z \int dk_z M(\vec{q},\vec{k}, k_{ez}) \nonumber\\
&\times& M^*(\vec{q},\vec{k}^\prime, k_{ez}) I(t,\omega, \hat{k}_e; k_z,k^\prime_z)
\eea
with
\bea
I(t,\omega, \hat{k}_e; k_z,k^\prime_z) & \equiv & -i \int_{t_0} ^{t} dt^{\prime\prime} \int_{t_0} ^{t} dt^\prime s(t^{\prime\prime}) s(t^\prime) e^{i\omega(t^{\prime\prime}- t^\prime)} \nonumber\\
&\times&G_{\vec{k}, \vec{k}^\prime}^< (t^\prime,t^{\prime\prime}).
\label{eq: idef}
\eea
Here $\vec{k} = ({\vec{k}}_{\parallel},k_z)$, $\vec{k}\,' = ({\vec{k}}_{\parallel},k^\prime_z)$ , and $G^<$ is the well known two-time (nonequilibrium) lesser Green's function~\cite{baym-kadanoff} given by
\bea
G_{\vec{k},\vec{k}^\prime}^< (t^\prime,t^{\prime\prime}) & = & i \sum_n \rho_n \langle {\Psi}_n |U(-\infty, t^{\prime\prime}) c_{\vec{k}\,'}^{\dag} U(t^{\prime\prime},t^\prime) \nonumber\\
&\times& c_{\vec{k}} U(t^\prime, -\infty)| \Psi_n\rangle \\
&\equiv& i \mathcal{Z}^{-1}{\rm Tr}[e^{-\mathcal{H}/(k_B T)} c_{\vec{k}\,'}^{\dag} (t^{\prime\prime})  c_{\vec{k}}(t^\prime)]
\eea
where $c_{\vec{k}\,'}^{\dag} (t^{\prime\prime})$ and $c_{\vec{k}}(t^\prime)$ are the electron creation and destruction operators in the Heisenberg picture appropriate to $\mathcal{H}_{\rm pump}(t)$:
\bea
c_{\vec{k}\,'}^{\dag} (t^{\prime\prime}) &\equiv& U(-\infty, t^{\prime\prime}) c_{\vec{k}\,'}^{\dag} U(t^{\prime\prime}, -\infty); \\
c_{\vec{k}} (t^\prime) &\equiv& U(-\infty, t^\prime) c_{\vec{k}} U(t^\prime, -\infty).
\eea
In the pumped case, this needs to be calculated using nonequilibrium Keldysh contour-ordered Green's function techniques~\cite{Keldysh} on the Kadanoff-Baym-Keldysh contour in the complex time plane.  Of course, if the pump pulse and the envelope function $s(t)$ vanish for $t>t_1$, then $P(t)$ becomes independent of time for $t>t_1$.
%Since the detectors continue to collect photoelectrons even after the probe is turned off, whence $I$ and $P$ become independent of $t$, $\lim_{t\rightarrow\infty} P(t,\omega, \hat{k}_e)$ is proportional to the total number of electrons collected for an energy and momentum-resolved experiment; for a three dimensional material, an integration over $k_z$ is required, while an integral over $\hat{k}_e$ is also required for a photoemission experiment that collects all electrons without angle resolution. Note also that we can let $t_0 \rightarrow -\infty$ as $s(t^\prime)$ keeps track of the turning on of the probe pulse.

\begin{figure}[th]
\centerline{\includegraphics [width=3.3in, angle=0, clip=on]  {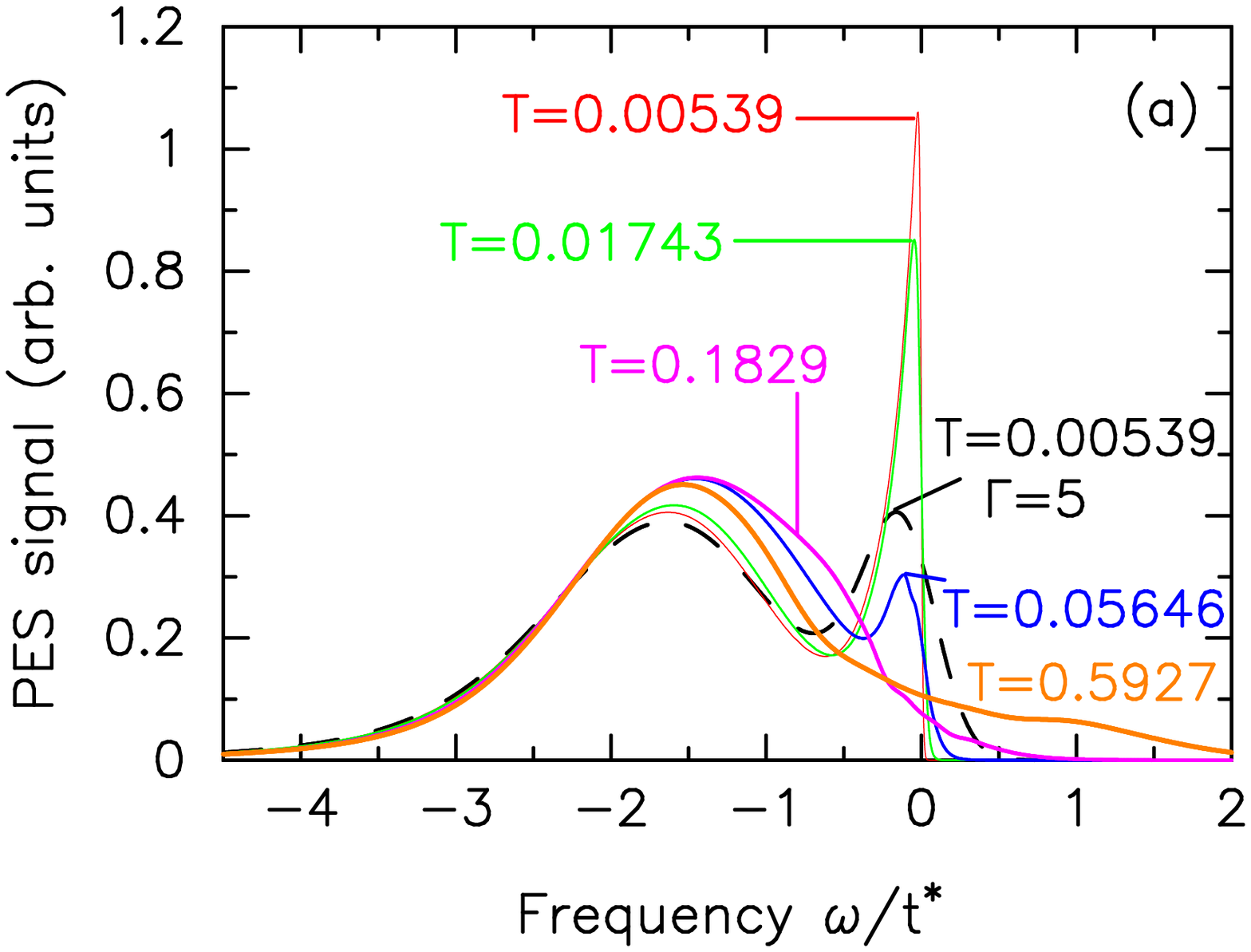}}
\centerline{\includegraphics [width=3.3in, angle=0, clip=on]  {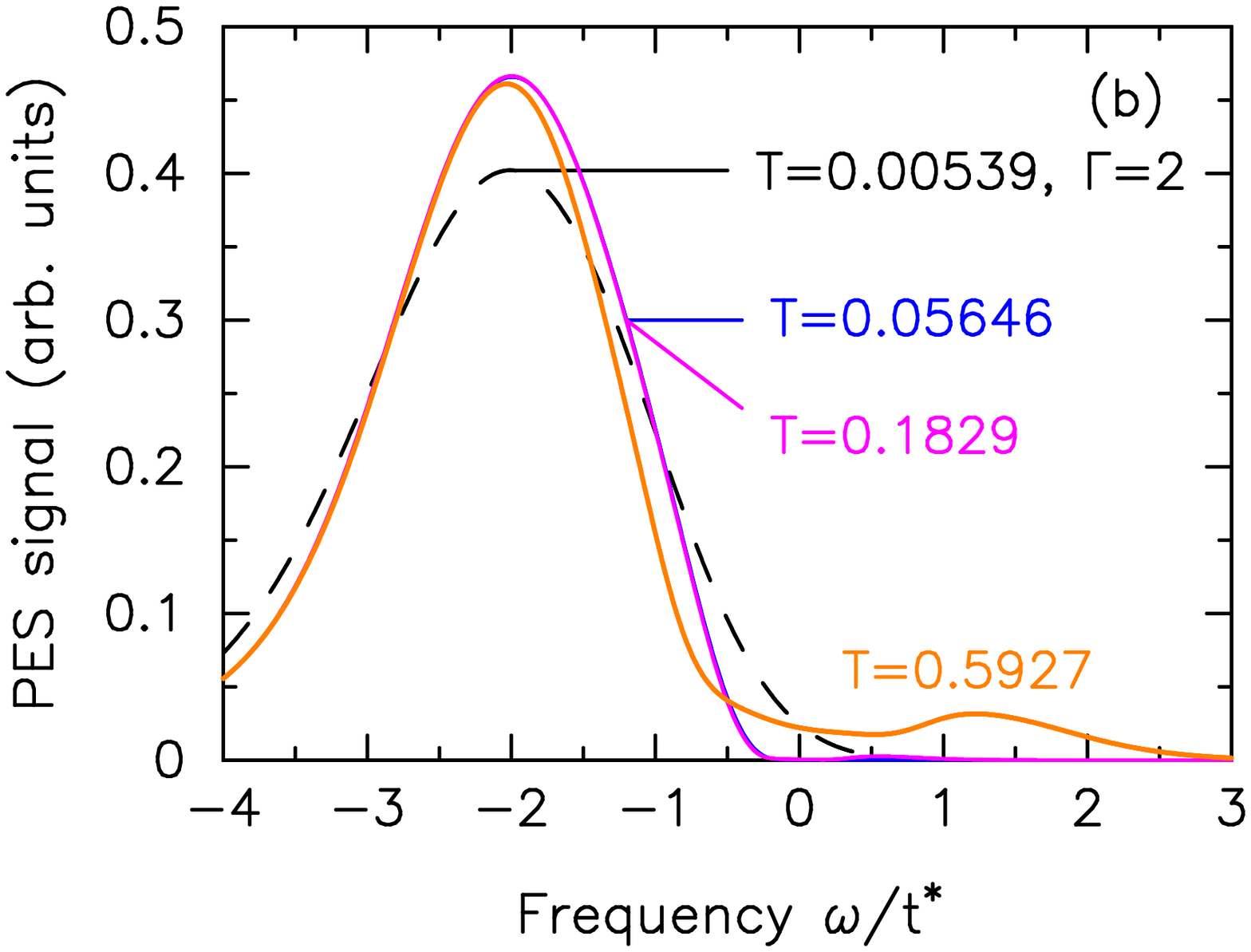}}
\caption[]{
(Color online) Time-resolved photoemission spectra for the Hubbard model at half filling and $U=3$ (top, width is $5$) or $U=4.2$ (bottom, width is $2$) at temperatures $T=0.00539$  (dashed line) versus the equilibrium continuous beam PES at different $T$s.  Note how the windowing effect is different from thermal broadening effects, especially for positive energies. The two low-temperature curves in panel (b) are almost overlapping except at positive frequencies, where there is a small difference.
}
\label{fig: window}
\end{figure}

In the equilibrium case, when $s(t) = 1$, $G^<_{\vec{k}}(t^\prime,t^{\prime\prime})$ is only a function of  $(t^\prime-t^{\prime\prime})$. Furthermore, in a highly anisotropic layered system, its $k_z$ dependence can be neglected. Then the ARPES transition rate (transition probability per unit time, which is what is relevant as the continuous beam probe pulse photoemits electrons at all times at a constant rate) is proportional to
\beq
\lim_{t \rightarrow \infty}\lim_{t_0\rightarrow -\infty} \frac {I(t,\omega, \hat{k}_e)}{(t-t_0)} =  -i G_{{\vec{k}}_{\parallel}}^<(\omega) = A_{{\vec{k}}_{\parallel}}(\omega)f(\omega)
\eeq
which is the  standard result.

In our numerical results, we focus on a material that can be well approximated by the $d \rightarrow \infty$ Hubbard model, for which the DMFT is exact~\cite{dmft}. The Hamiltonian is
\begin{equation}
 \mathcal{H}=-\frac{t^*}{2\sqrt{d}}\sum_{\langle i,j\rangle \sigma}c^\dagger_{i\sigma}c^{}_{j\sigma}+U\sum_i
c^\dagger_{i\uparrow}c^{}_{i\uparrow}c^\dagger_{i\downarrow}c^{}_{i\downarrow}.
\end{equation}
Here electrons hop between nearest neighbor sites on an infinite dimensional hypercubic lattice (leading to a Gaussian noninteracting density of states) and they have an opposite spin repulsive interaction $U$ when two electrons are on the same site. In the rest of our paper, all energies are in units of $t^*$. We consider only TR-PES, where we integrate over the direction of the photoemitted electron.  For simplicity we neglect the dependence of the matrix element $M$ on momenta~\cite{mat-el} and treat it as a constant, and we ignore all $k_z$ dependence in $I$. We also assume that the pump pulse acts solely to heat the electronic system which rapidly thermalizes into a quasi-equilibrium distribution at an effective electron temperature $T_{el}$,  so that we can focus on the effect of the windowing due to the probe pulse envelope $s(t)$.

We solve the DMFT problem using the numerical renormalization group~\cite{NRG} for the retarded Green's function.  We take $\Lambda=1.8$ and keep 800 states per iteration on the Wilson chain. Once the retarded Green's function is found, we multiply the imaginary part by $-2if(w)$ to get the lesser Green's function, then we Fourier transform to real time to find $G^<(t-t^\prime)$.  Next we evaluate Eq.~(\ref{eq: idef}), which is proportional to the TR-PES spectra given the above assumptions.  We take the shape function $s(t)$ to be a Gaussian $s(t)=\exp[-(t-\bar t)^2/\Gamma^2]/(\Gamma\sqrt{\pi})$ with a varying width $\Gamma$  and a peak time~\cite{tbar} $\bar t$;  We work at half filling, with various $U$ values to examine both metallic and insulating phases. Our PES signals are normalized so that the integrated weight in all spectra are identical.

In Fig.~\ref{fig: PES_U}, we compare the continuous beam PES [$s(t)= 1$] with the pump-probe PES for  Gaussian probe shape functions of varying width. Panels (a) and (b), with  $U=3$,  correspond to a strongly correlated metal, and panel (c), with $U=4.2$, to a Mott insulator.  While the broader higher energy features of the spectral function are determined accurately once the width is on the order of $\Gamma=5$, sharp features like the quasiparticle peak, or kink-like features at higher temperatures (see inset at frequencies near $\pm 0.2$), require much broader windows to be accurately captured.  We see similar results for the Mott insulating phase, where the spectra have no sharp features, except for some positive frequency features due to the thermal excitations at high $T$ as in the inset; results at lower $T$ are essentially the same, but have no  thermally activated peaks at positive frequency.

It is interesting to ask whether the broadening effect of the finite probe pulse width can be mimicked by a finite temperature thermal broadening.  We examine this in Fig.~\ref{fig: window}, where we compare the metallic PES (top panel) for $T=0.00539$ and a probe pulse width of $\Gamma=5$ (dashed line) with the continuous beam PES for various temperatures.  We see the spectral shapes are quite different.  The reason is that the high temperatures needed to broaden the quasiparticle peak to the same extent as obtained from the windowing effect of the probe pulse also generate higher energy upper Hubbard band contributions. The windowing effect alone just does not have these. This is further supported by the results for the insulating phase (lower panel), where we see similar effects.

In conclusion, we have developed a full many-body theory under the conventional photoemission assumptions for pump-probe time resolved photoemission that takes into account two new effects: (1) the nonequilibrium dynamics brought on by the large fields in the pump pulse and (2) the windowing effect due to the finite width of the probe pulse.  We find that the PES signal including all the nonequilibrium effects can be represented in terms of integrals of the lesser Green's function in the presence of the pump pulse, and that the windowing effect can make it difficult to extract narrow energy features unless the pulse width is wide enough in the time domain.  This latter effect is quite different from the effect of raising the temperature, which also causes broadening, but of a qualitatively different character. Given the experimental parameters and lowest-energy bandwidth of 1T-TaS$_2$, it is possible that the windowing effect could be playing a role in that system.

{\it Acknowledgments:}  J. K. F. acknowledges support from the National Science Foundation under grant number DMR-0705266.
H. R. K. is supported under ARO Award W911NF0710576 with funds from the DARPA OLE Program.
Th. P. acknowledges support from the collaborative research center (SFB) 602.
We also acknowledge useful discussions with
T. Devereaux,
M. Grioni,
B. Moritz,
L. Perfetti, and
K. Rossnagel.


\begin{thebibliography}{99}
\bibitem{tbar}
In real pump-probe experiments, $\bar t$ determines the time delay, and hence the effective electron temperature $T_{el}$ in a quasiequilibrium description, but thereafter plays no further role. $I$  can then also be expressed as the frequency domain convolution
$\lim_{t \rightarrow \infty}\lim_{t_0\rightarrow -\infty} I(t,\omega, \hat{k}_e) =  -i\int d\nu G_{{\vec{k}}_{\parallel}}^<(\omega-\nu)) \left |{\tilde s}(\nu)\right |^2/2\pi$
where ${\tilde s}(\nu)$ is the Fourier transform of $s(t^\prime)$.
\end{thebibliography}
\end{document}